\documentclass[letterpaper]{article} 
\usepackage{aaai25}  
\usepackage{times}  
\usepackage{helvet}  
\usepackage{courier}  
\usepackage[hyphens]{url}  
\usepackage{graphicx} 
\usepackage{natbib}  
\usepackage{caption} 
\usepackage{algorithm}
\usepackage{algorithmic}
\usepackage{multirow}

\usepackage{newfloat}
\usepackage{listings}
\DeclareCaptionStyle{ruled}{labelfont=normalfont,labelsep=colon,strut=off} 
\floatstyle{ruled}
\newfloat{listing}{tb}{lst}{}
\floatname{listing}{Listing}
\usepackage{amsmath}
\usepackage{amsfonts}
\usepackage{subfigure}
\usepackage{color}
\usepackage{enumitem}

\definecolor{dkgreen}{rgb}{0,0.6,0}
\definecolor{gray}{rgb}{0.5,0.5,0.5}
\definecolor{mauve}{rgb}{0.58,0,0.82}

\title{Beyond Local Views: Global State Inference with Diffusion Models\\for Cooperative Multi-Agent Reinforcement Learning}
\author{
    Zhiwei Xu\textsuperscript{\rm 1}, Hangyu Mao\textsuperscript{\rm 2}, Nianmin Zhang\textsuperscript{\rm 3}, Xin Xin\textsuperscript{\rm 1}, Pengjie Ren\textsuperscript{\rm 1}, Dapeng Li\textsuperscript{\rm 4}, \\Bin Zhang\textsuperscript{\rm 4}, Guoliang Fan\textsuperscript{\rm 4}, Zhumin Chen\textsuperscript{\rm 1}, Changwei Wang\textsuperscript{\rm 5}, Jiangjin Yin\textsuperscript{\rm 6}
}
\affiliations{
    \textsuperscript{\rm 1}Shandong University \quad \textsuperscript{\rm 2}Kuaishou Technology\\
    \textsuperscript{\rm 3}University of Hong Kong \quad \textsuperscript{\rm 4}Institute of Automation, Chinese Academy of Sciences\\    
    \textsuperscript{\rm 5}Qilu University of Technology \quad \textsuperscript{\rm 6}Huazhong Agricultural University\\
    zhiwei\_xu@sdu.edu.cn

}

\usepackage{bibentry}

\begin{document}

\maketitle

\begin{abstract}
In partially observable multi-agent systems, agents typically only have access to local observations. This severely hinders their ability to make precise decisions, particularly during decentralized execution. To alleviate this problem and inspired by image outpainting, we propose State Inference with Diffusion Models (SIDIFF), which uses diffusion models to reconstruct the original global state based solely on local observations. SIDIFF consists of a state generator and a state extractor, which allow agents to choose suitable actions by considering both the reconstructed global state and local observations. In addition, SIDIFF can be effortlessly incorporated into current multi-agent reinforcement learning algorithms to improve their performance. Finally, we evaluated SIDIFF on different experimental platforms, including Multi-Agent Battle City (MABC), a novel and flexible multi-agent reinforcement learning environment we developed. SIDIFF achieved desirable results and outperformed other popular algorithms.
\end{abstract}

%

\section{Introduction}

As cooperative multi-agent reinforcement learning (MARL) advances, many proposed methods are being successfully applied to a variety of practical problems, such as online ride-hailing platforms~\cite{LiQJYWWWY19}, drone swarm management~\cite{WangYMXWWG23}, and energy system scheduling~\cite{ZHANG202370}.
While some methods rely on explicit communication mechanisms~\cite{DasGRBPRP19, LiXZZZF24, JiangL18}, most algorithms follow the centralized training with decentralized execution (CTDE) framework~\cite{LoweWTHAM17}. 
Agents can share information or access the global state in the training phase of MARL algorithms that adhere to the CTDE paradigm. During the decentralized decision-making phase, agents must select appropriate actions based solely on their individual local observations. This limitation is especially pronounced in partially observable Markov decision processes (POMDPs)~\cite{Spaan12}, where the absence of global state awareness during execution can impede the agents' ability to make optimal choices. \looseness=-1

\begin{figure}[t]
    \centering
    \includegraphics[width=3.0 in]{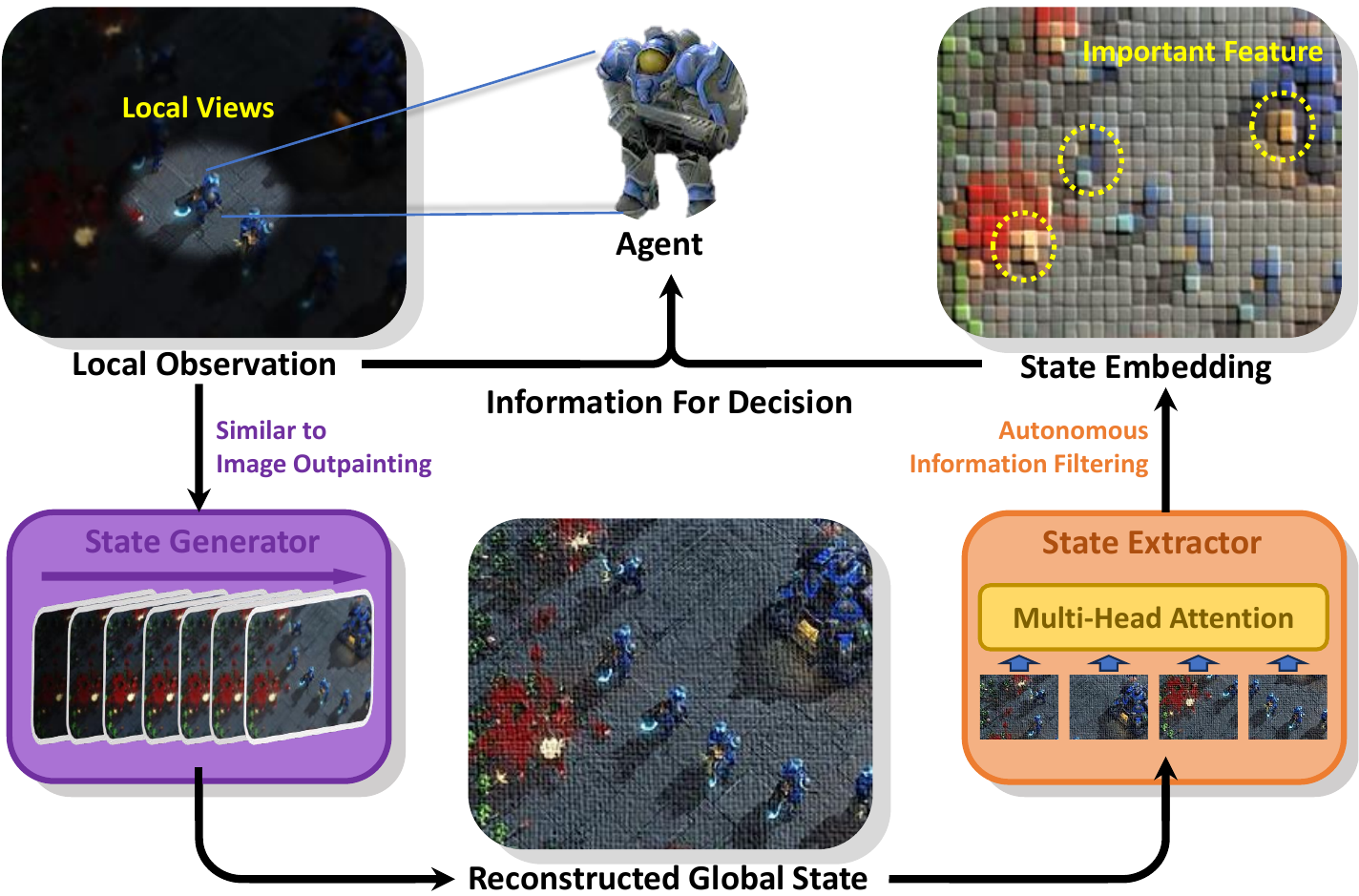}
    \caption{In the multi-agent system, an agent under the SIDIFF framework chooses reasonable actions after two steps: 1) reconstructing the state and 2) extracting information.}
    \label{fig:intro}
\vspace{-0.1 in}
\end{figure}

To address the challenges of partial observability, some prevalent methods~\cite{HausknechtS15, RMADDPG} involve integrating recurrent neural networks (RNNs)~\cite{RNN} within reinforcement learning models.
By incorporating historical information into their decision-making process, agents can mitigate issues arising from incomplete information. However, these approaches completely disregard the utilization of any additional information about global state.  
An alternative approach~\cite{SIDE, GCN} concentrates on forecasting a low-dimensional, abstract representation of the global state. This method employs representation learning techniques, such as contrastive learning~\cite{ChenK0H20}, to distill global state information. This paradigm avoids the complexities involved in directly reconstructing the global state.
However, since the processes of representation learning and reinforcement learning are relatively independent, there is no guarantee that the derived additional information will enhance the reinforcement learning process. In summary, neither of the methods above effectively leverages the global state during the decentralized execution phase.

Decision-making based on reconstructed global states has a biological foundation. 
In practical scenarios, humans frequently make decisions with incomplete information by inferring the broader context from local cues~\cite{Simon1955ABM, Friston2010TheFP}. Thoughtful and rational decisions can only be made when both the inferred global state and the immediate local observations are considered.
This principle applies equally to cooperative multi-agent systems. Suppose each agent can infer the current global state in a distributed manner. In that case, it not only improves the decision-making process but also empowers the agent to select useful information autonomously. This approach ensures that global state information is utilized comprehensively and efficiently. 

However, reconstructing the global state from local observations within the original state space remains a formidable challenge. Some prior studies~\cite{HuangSZC20, HanDT20} have explored traditional generative models, such as Generative Adversarial Networks (GANs)~\cite{Goodfellow2022GenerativeAN} and Variational Autoencoders (VAEs)~\cite{KingmaW13}, for reconstructing the global state. Nevertheless, they have typically been tailored to simple single-agent tasks or multi-agent scenarios with low-dimensional state spaces. Furthermore, other works employing particle filter methods to build world models have encountered similar drawbacks.

To overcome this challenge, we proposed the \textbf{S}tate \textbf{I}nference with \textbf{DIFF}usion Models (\textbf{SIDIFF}) framework.
Diffusion models can extend the boundaries of an existing image, a technique known as \emph{image outpainting}~\cite{Bertalmio2000image}, which has been successfully applied in well-known computer vision projects like DALL-E~\cite{RameshPGGVRCS21}.
Inspired by the diffusion model, we use it as a \emph{state generator} for agents to infer the original global state in a distributed manner based on local observations. We then introduce the Vision Transformer (ViT)~\cite{DosovitskiyB0WZ21} architecture as a \emph{state extractor} to extract information from the reconstructed global state effectively. With access to comprehensive global state information, agents can improve training efficiency in partially observable online cooperative reinforcement learning tasks. 
Therefore, the agent can reconstruct the global state in the original state space, and then refines this reconstructed state to yield more efficient global information, as illustrated in Figure~\ref{fig:intro}. As a result, during the decentralized execution phase, the agent in the SIDIFF framework can select appropriate actions leveraging both global insights and local observations. 
SIDIFF has achieved good performance in recent popular partially observable benchmarks, as well as in the new environment we proposed, \textbf{M}ulti-\textbf{A}gent \textbf{B}attle \textbf{C}ity (\textbf{MABC}). To the best of our knowledge, SIDIFF is the first framework that uses diffusion models for reconstructing the global state in online multi-agent reinforcement learning tasks.\looseness=-1


\section{Related Work}

\subsection{Partially Observable Problems}

In both single-agent and multi-agent scenarios, the partially observable Markov decision process (POMDP)~\cite{POMDP} requires agents to make decisions based on incomplete information. Due to the possibility of a lack of critical information, agents are frequently limited to selecting locally optimal actions. Therefore, bridging this information gap is an important research topic in reinforcement learning. 
The most popular idea is to employ recurrent neural networks (RNNs) to integrate local observation data over time, thereby endowing agents with a form of long-term memory. As a seminal work in this field, DRQN~\cite{HausknechtS15} has influenced various MARL methods, including R-MADDPG~\cite{RMADDPG}, QMIX~\cite{QMIX}, and MAPPO~\cite{MAPPO}. 
Another popular technique is belief tracking with particle filters~\cite{MaKHLY20, MaKHL20}, which typically requires a predefined model. Some research~\cite{HuangSZC20, HanDT20, HausknechtS15, IglZLWW18} has focused on generating latent state representations through variational inference. Variational autoencoders (VAE)~\cite{KingmaW13}, as generative models, can compress complex observations to a compact latent space by maximizing the evidence lower bound. However, these methods focus primarily on the issue of inaccessible global state during training, often overlooking the potential utility of states. Communication~\cite{FoersterAFW16, PengYWYTLW17} emerges as a distinct method to mitigate the need for global information. Agents in communication methods exchange information, which is either manually specified or learned. However, this approach also incurs high communication costs. 
Additionally, some studies~\cite{0001WSZ21, 00050ZZBF22, PasztorKB23} model opponents or the environment as auxiliary tasks to predict the behavior of other entities. However, these methods are unsuitable for complex environments and may be constrained by issues such as compounding errors.

\subsection{Diffusion Models for Reinforcement Learning}

Diffusion models~\cite{HoJA20}, renowned for their capacity to generate diverse data and capture multimodal distributions, are increasingly being integrated into reinforcement learning to improve performance and sample efficiency. The most mainstream method is to fit the dynamics of the environment with diffusion models, which serve as a planner~\cite{BrehmerBHC23, JannerDTL22, LiangMDNTL23}. Guided by objectives such as expected return-to-go, diffusion models can generate trajectories that align with both the given directives and the environmental dynamics. Moreover, just as diffusion models are used for data augmentation in computer vision, they also hold promise as data synthesizers in offline reinforcement learning~\cite{ChenK0K23, LuBTP23}. Some studies~\cite{AdaOU24, ChiFDXCBS23, IDQL} have employed diffusion models directly as policies, addressing challenges such as over-conservatism and limited adaptability to diverse datasets in offline settings.
Furthermore, a few works also applied diffusion models in fields such as quality diversity reinforcement learning and multi-agent cooperation tasks. However, almost all of the above-mentioned studies combined diffusion models and reinforcement learning algorithms in offline settings or within single-agent tasks, primarily to mitigate common offline issues. In contrast, the SIDIFF framework we propose is intended for online multi-agent tasks, explicitly targeting the challenges of partial observability. Regardless of the background or concerns, SIDIFF is entirely orthogonal to previous work.


\section{Preliminaries}

\subsection{Dec-POMDPs}

The fully cooperative multi-agent decision-making problem can be modeled as a decentralized partially observable Markov decision process (Dec-POMDP). Within this problem, there are $n$ agents denoted by the set $A = \{1,\dots,n\}$. 
$s\in S$ is the global state of the environment. The observation function, denoted by $O:A\times S \to Z$, assigns distinct local observations to each agent based on the current global state $s$. 
At each time step, each agent $a$ selects an action $u\in U$ based on its local observation $z=O(a, s)$. 
$\boldsymbol{u}\in \boldsymbol{U} \equiv U^n$ represents the joint action of all agents. In the Dec-POMDP problem, all agents share a reward function $r(s,\boldsymbol{u}):S \times U \to \mathbb{R}$, which ensures full cooperation between agents when making decisions. 
$p(s^{\prime} \mid s, \boldsymbol{u}):S \times \boldsymbol{U} \times S \to [0,1]$ denotes the transition function of the system. The goal of all agents is to maximize the discounted return $\sum_{i=0}^\infty \gamma^i r_{t+i}$, where $\gamma$ is the discount factor.

\subsection{Denoising Diffusion Probabilistic Models}

The diffusion model draws inspiration from nonequilibrium thermodynamics~\cite{Sohl-DicksteinW15}. It encompasses two key processes: the forward and the reverse processes. Noise is gradually introduced to the original data $x$ during the forward process. Conversely, the reverse process systematically removes this noise, thereby restoring the data to its initial state. First, sample $x_0 \sim q(x_0)$ from the actual data distribution. The forward process sequentially introduces Gaussian noise to the data according to the predefined variance parameter $\beta_{1:K}$. Specifically, the $x_k$ obtained in the $k$-th iteration of the forward process is derived as follows:\looseness=-1
\begin{equation*}
    q(x_k \mid x_{k-1}) = \mathcal{N}(x_k ; \sqrt{1-\beta_k}x_{k-1}, \beta_k \boldsymbol{I} ).
\end{equation*}
When deriving $x_k$ during the forward process, it can be directly computed from the initial data $x_0$ without the necessity of sequentially calculating the intermediate steps $x_{1:k-1} = \{x_1, x_2, \dots, x_{k-1}\}$. The specific calculation formula can be written as follows:
\begin{equation*}
    x_k=\sqrt{\bar{\alpha}_k} x_0+\sqrt{1-\bar{\alpha}_k} \epsilon\left(x_k, k\right),
\end{equation*}
where $\bar{\alpha}_k = \prod_{i=1}^k \alpha_i$ and $\epsilon\left(x_k, k\right) \sim \mathcal{N}(\boldsymbol{0}, \boldsymbol{I})$, with $\alpha_k = 1 - \beta_k$. In the reverse process, the diffusion model is trained to learn a conditional distribution that iteratively denoises and restores the original data. The formula to recover $x_{k-1}$ from $x_k$ is as follows:
\begin{equation}
    p(x_{k-1} \mid x_{k}) = \mathcal{N}(x_{k-1} ; \mu(x_k,k), \sigma(x_k,k) ).
\label{eq:diff_recover}
\end{equation}
And $\mu$ and $\sigma^2$ can be given as follows:
\begin{equation*}
\begin{aligned}
\mu\left(x_k, k\right)&=\frac{1-\bar{\alpha}_{k-1}}{1-\bar{\alpha}_k} \sqrt{\alpha_k} x_k+\frac{\sqrt{\bar{\alpha}_{k-1}} \beta_k}{1-\bar{\alpha}_k} x_0\\&=\frac{1}{\sqrt{\alpha_k}}\left(x_k-\frac{\beta_k}{\sqrt{1-\bar{\alpha}_k}} \epsilon\left(x_k, k\right)\right),\\
\sigma^2\left(x_k, k\right)&=\beta_k \frac{1-\bar{\alpha}_{k-1}}{1-\bar{\alpha}_k},
\end{aligned}
\end{equation*}
where $\sigma^2$ is a constant value, and $\mu$ is a function dependent on $\epsilon(x_k, k)$. We employ a neural network to estimate $\epsilon(x_k, k)$, denoted as $\epsilon_\theta(x_k, k)$, where $\theta$ represents the parameters of the neural network. The diffusion model is trained by minimizing the following loss function:
\begin{equation*}
    \mathcal{L}:=\mathbb{E}_{k,x_0}\left[\left\|\epsilon_k-\epsilon_\theta\left(\sqrt{\bar{\alpha}_k} x_k+\left(\sqrt{1-\bar{\alpha}_k}\right) \epsilon_k, k\right)\right\|^2\right].
\end{equation*}
$\epsilon_k$ is the actual noise sampled during the forward process.


\section{Methodology}

The core idea behind SIDIFF is to empower the agent to deduce the global state based on local observations during decentralized execution, leveraging this inferred information to refine their decision-making processes. In this section, we delve into the architecture of SIDIFF, which is composed of two key components: the \emph{State Generator} and the \emph{State Extractor}. 
For ease of exposition, when variables in this paper are subscripted with two indices, the first index refers to the specific time step within an episode, while the second indicates the iteration of the diffusion process. 
For instance, $\hat{s}_{t,k}$ represents the state $s$ generated by the diffusion model at time step $t$ in the $k$-th iteration of the diffusion process.

\begin{figure}[!t]
    \centering
    \includegraphics[width=3.3 in]{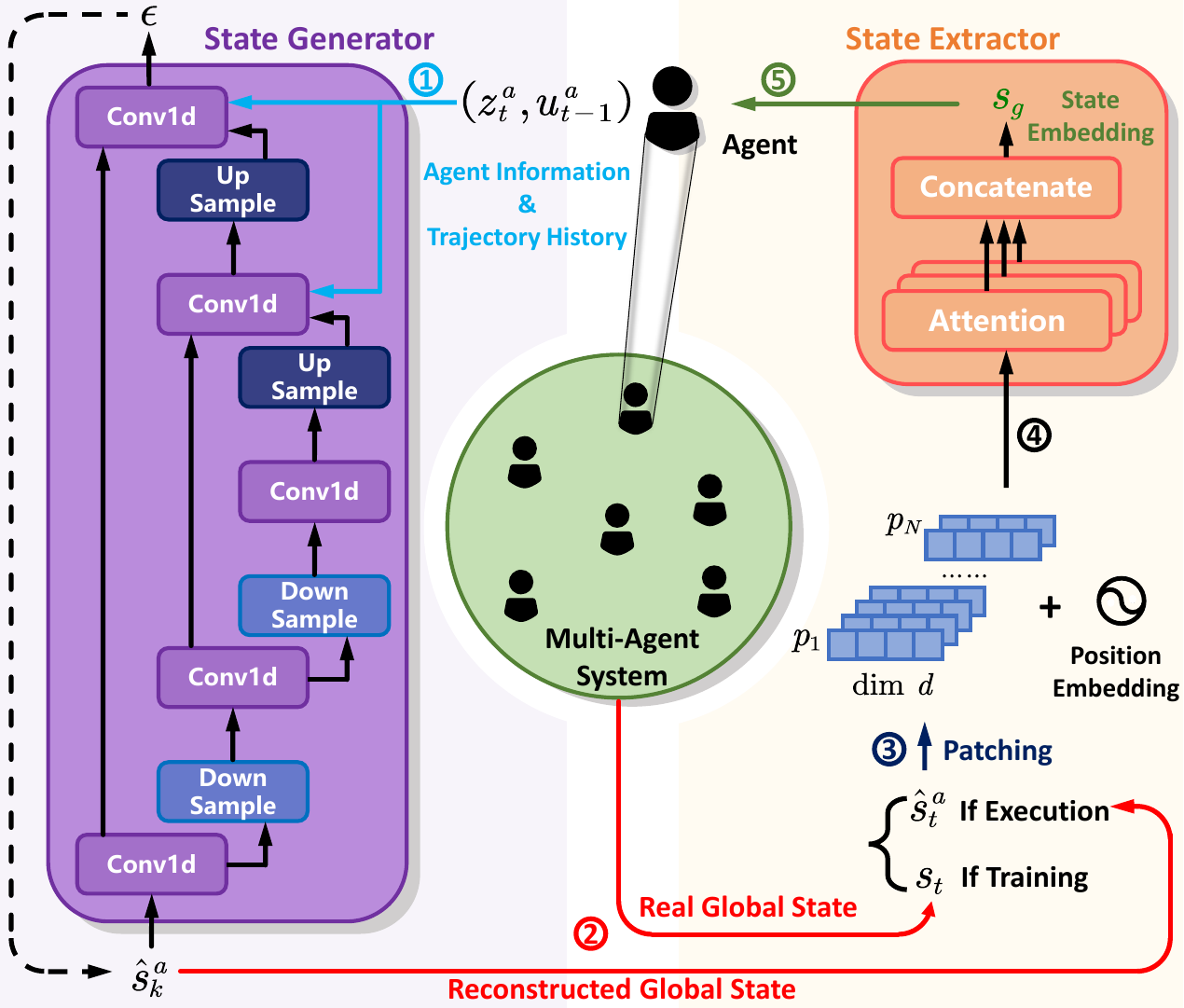}
    \caption{The overall framework and workflow of SIDIFF.}
    \label{fig:framework}
\end{figure}

\subsection{State Generator}

First, we introduce the state generator module, which models data distribution using the diffusion model. The state generator is intended to infer the global state automatically, similar to image outpainting tasks.
The state generator is patterned after the U-Net architecture~\cite{RonnebergerFB15}, as depicted in Figure~\ref{fig:framework}. It comprises a series of one-dimensional convolutional residual blocks. The structure of the state generator is divided into two symmetrical segments: an encoder and a decoder. The encoder extracts contextual state information, while the decoder reconstructs the original state.
Moreover, the decoder employs skip-connection techniques to merge feature maps from the encoder with those in the decoder, facilitating the integration of multi-scale features. Similar to the traditional diffusion model, the state generator takes the reconstructing state as input and outputs the predicted noise $\epsilon$. Further details will be given below.

The state generator needs to reconstruct the global state based on local observations of individual agents. However, in partially observable problems, agents may encounter the challenge of having identical local observations, even in fundamentally distinct global states. Therefore, to ensure the uniqueness of these conditions used to generate the state as much as possible, we integrate time-sensitive data into decoders.
SIDIFF involves the current diffusion iteration $k$, along with unique identifiers $a$ for each agent, and the trajectory history $\tau$. These elements are employed as conditions to predict the noise. The embeddings $e_K$, $e_\text{A}$, and $e_\mathcal{T}$ are the features derived from the affine transformation of the aforementioned conditions.
Assuming $f_L$ represents the output from the preceding layer within a particular decoder block, and $f_E$ denotes the skip-connected feature from the corresponding encoder block, the final input for the decoder block would be written in the form:
\begin{equation*}
    \operatorname{CONCAT}\Big(\big(e_\mathcal{T}+e_\text{A}\big) \cdot f_L + e_K, \  f_E\Big).
\end{equation*}
The trajectory history of an agent is essential for the state generator to reconstruct the global state. In addition, $e_K$ plays a pivotal role in the training and sampling phases of the diffusion model. Like other classic works, we incorporate it as a bias term. Finally, these conditional variables are concatenated with the output of the symmetric encoder block to form the input of the decoder block. Referring to classifier-free diffusion models~\cite{CLFEDM}, we jointly train the unconditional and conditional models simply by randomly setting the information of these agents to the unconditional class identifier $\emptyset$ with some probability $\beta\sim\operatorname{Bern}(p)$. In practical implementation, the unconditional class identifier $\emptyset$ is typically substituted with a vector of zeros.

The training and sampling regimen of the state generator closely resembles that of traditional diffusion models. We set the diffusion process duration to $K\in \mathbb{N}^{+}$. The goal of training is for the state generator to combine the iteration $k$, the local observation of the agent, and the ground-truth global state to predict the noise $\epsilon_k$ introduced in the forward process, rather than directly reconstructing the original global state as the VAE does. The loss function for training the state generator is as follows:
\begin{equation}
\begin{aligned}
    &\mathcal{L}_{SG}(\theta)= \\&\mathbb{E}_{k, s_{t,0} \in \mathcal{D}, \beta \sim \operatorname{Bern}(p)}\left[\big\|\epsilon_k-\epsilon_\theta\left(\hat{s}_{t,k},(1-\beta) \boldsymbol{c}+\beta \emptyset, k\right)\big\|^2\right],
\end{aligned}
\label{eq:diff_loss}
\end{equation}
where $\theta$ denotes the network parameters of the state generator, and $\boldsymbol{c} = \{a,\tau\}$ is the conditional dependencies required for generating the global state. Even in a fully distributed framework, agents can acquire these conditions. In the inference phase, each agent initializes $\hat{s}_{t,K} \sim \mathcal{N}(\boldsymbol{0}, \boldsymbol{I})$ and then removes the noise $\epsilon$ predicted by the state generator to obtain the denoised state. Precisely, the inferred global state at the $k$-th iteration of the reverse process can be calculated:
\begin{equation}
    \hat{s}_{t,k-1}=\frac{1}{\sqrt{\alpha_k}}\left(\hat{s}_{t,k}-\frac{1-\alpha_k}{\sqrt{1-\bar{\alpha}_k}} \boldsymbol{\epsilon}_\theta\left(\hat{s}_{t,k}, \boldsymbol{c}, k\right)\right)+\sigma_k \mathbf{z},
\label{eq:denoise}
\end{equation}
where $\mathbf{z} \sim \mathcal{N}(\boldsymbol{0}, \boldsymbol{I})$. This iterative procedure continues until the original global state $\hat{s}_{t,0}$ is restored. The incorporation of the state generator module endows the agent with the ability to conduct distributed inference of the original global state.

\subsection{State Extractor}

On the one hand, the global state encompasses critical information that agents may depend on for decision-making, which is frequently unavailable through local observations alone. On the other hand, the global state contains a large amount of redundant information irrelevant to the decision-making of individual agents.
Directly incorporating the global state into the agent network may place an undue burden on individual agents, potentially reducing learning efficiency. This phenomenon has been well-documented in some prior research~\cite{LiXZZZF24, GuanCYWYZ022}. Consequently, extracting decision-relevant information from the inferred global state $\hat{s}$ is also an important contribution of SIDIFF. In the following, we explain how to extract crucial decision-relevant information from the original global state.\looseness=-1

The global state is defined as containing all information in a multi-agent system. However, for an individual agent, only a small portion of the global state is actually useful for decision-making. Therefore, it is imperative to distill the global state into a meaningful abstraction, a task that typically demands expert knowledge. 
Drawing inspiration from the Vision Transformer (ViT) in computer vision, we first divide the global state into a series of fixed-size 1D patches $\boldsymbol{P} = \{p_i\}$, $i\in\{1,\dots,N\}$. These vectors are then arranged in sequence to form an ordered sequence. Simultaneously, position embeddings are integrated to infuse positional information into the patch vectors, preserving the intrinsic semantics of the original global state. These vectors serve as the input to the Transformer model. 
By utilizing the multi-head attention mechanism~\cite{Attention}, we derive the abstracted feature embedding $s_g$ for the global state. It can be computed by the following equation:
\begin{equation}
s_g=\operatorname{CONCAT}\Big(\operatorname{Softmax}\big(\frac{\boldsymbol{P}\boldsymbol{W}^Q_j(\boldsymbol{P}\boldsymbol{W}^K_j)^\top}{\sqrt{d}}\big)\boldsymbol{P}\boldsymbol{W}^V_j\Big),
\label{eq:abstracted_feat}
\end{equation}
where $j\in\{1,\dots,H\}$. $d$ is the scaling factor equal to the dimension of each patch, and $H$ is the number of attention heads. Finally, $s_g$ will be concatenated with the agent's local observation and fed into the agent network.

As illustrated in Figure~\ref{fig:framework}, the state extractor converts the original global state into information that is more conducive to the agent's decision-making and feeds it into the agent's model. 
Unlike other studies that replace the global state with low-dimensional embeddings, the state extractor is directly optimized end-to-end by the reinforcement learning process rather than multiple loss functions or multi-stage optimization. 
Therefore, when making decisions, the agent actively selects information from the global state conducive to its own decision-making, rather than passively absorbing information that other relatively independent modules have processed.

\begin{figure*}[!t]
    \centering
    \includegraphics[width=6.2 in]{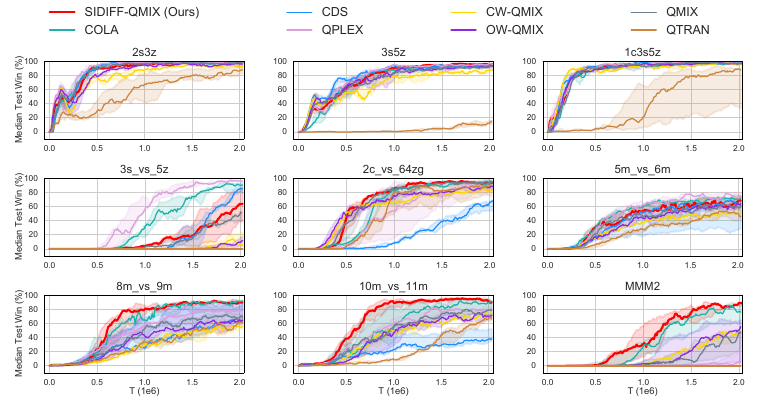}
    \vspace{-0.175 in}
    \caption{Performance comparison with baselines in SMAC. SIDIFF-QMIX outperforms QMIX in almost all scenarios.}
    \label{fig:smac_result}
    \vspace{-0.10 in}
\end{figure*}

\subsection{Centralized Training with Decentralized Execution}

Although diffusion models can generate more complex data compared to other generative models, it is achieved by breaking down the entire generation process into multiple steps, which incurs a higher cost in terms of wall-clock time. 
SIDIFF is an online multi-agent reinforcement learning framework, in contrast to previous research, which was primarily conducted offline and relied solely on predetermined datasets with no real-time environmental interaction. 
Therefore, to reduce the additional time cost introduced by the denoising process of the diffusion model, SIDIFF utilizes the ground-truth state $s$ during the training phase. 
Only during the decentralized decision-making phase does the agent use the state generator to produce $\hat{s}$ to replace the true global state $s$. 
Thus, the multi-step reverse process only impacts the evaluation phase, accounting for a small portion of the algorithm training process. This approach ensures that the time cost is not significantly increased while also maintaining the stability of the algorithm training.  
Moreover, in our implementation, we only update the state generator prior to evaluation, in accordance with Eq.~\eqref{eq:diff_loss}, which also reduces training time. Last but not least, more advanced diffusion model methods~\cite{DPM, SongD0S23} can be applied in the future to further enhance the efficiency of SIDIFF during the decentralized execution phase.

SIDIFF can be applied to various multi-agent reinforcement learning algorithms that adhere to the CTDE paradigm, including value decomposition and policy-based methods. 
The most significant modification in the SIDIFF-augmented variants of these algorithms is the introduction of $s_g$, which contains important information about the global state. Note that $s_g$ is generated by the state generator and subsequently abstracted by the state extractor. 
For value decomposition methods with SIDIFF, the value function for agent $a$ is defined by $Q_a(\tau^a, s_g)$. And we can obtain the joint value function $Q_\text{tot}$:
\begin{equation*}
    Q_\text{tot}=f_{M}\big(Q_a,a\in\mathcal{A}\big),
\end{equation*}
where the function $f_M$ represents the mixing network, which can take different forms depending on the specific value decomposition method. 
For policy-based methods like MAPPO, the Actor network is denoted as $\pi_n(\tau_n, s_g)$ in their SIDIFF-augmented versions. It can be seamlessly integrated into the policy gradient loss function of the original algorithm. The details for the variants with SIDIFF of these two different algorithms are presented in \textbf{Appendix~A}.


\section{Experiment}

To validate whether SIDIFF can be applied to various algorithms and enhance performance, we have obtained extensive experimental results on three different experimental platforms: SMAC~\cite{SamvelyanRWFNRH19}, VMAS~\cite{BettiniKBP22}, and MABC. Both SMAC and VMAS are well-known environments in the field of multi-agent systems, with partial observability and a diverse set of scenarios for evaluating cooperation between agents. Multi-Agent Battle City (MABC) is a novel experimental platform we have proposed, in which agents only have access to local observations and scenarios can be flexibly customized. We conducted a case study in a specific MABC scenario to demonstrate the importance of the global state in decision-making. Finally, through a series of ablation experiments and visualization, we show that both the state generator and state extractor in SIDIFF are indispensable. Detailed information about the testbeds can be found in \textbf{Appendix~B}.

\begin{figure*}[!t]
    \centering
    \includegraphics[width=6.5 in]{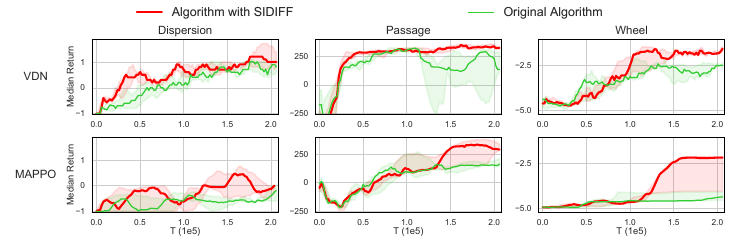}
    \vspace{-0.175 in}
    \caption{Comparison of our approach against baseline algorithms on Vectorized Multi-Agent Simulator.}
    \label{fig:vams_result}
    \vspace{-0.10 in}
\end{figure*}

\subsection{StarCraft Multi-Agent Challenge}

SMAC is a cooperative multi-agent reinforcement learning environment based on the real-time strategy game StarCraft II. It includes a variety of unique scenarios in which agents select actions based on local observations within their visual field. The goal of all scenarios is to command allied units to eliminate enemy units, which are controlled by built-in heuristic algorithms. 
We applied SIDIFF to the value decomposition method QMIX and compared it with other baselines such as CDS~\cite{LiWWZYZ21}, QPLEX~\cite{WangRLYZ21}, Weight-QMIX, QTRAN~\cite{RashidFPW20}, and the original QMIX. To ensure fairness, all experiments were conducted with the same environment settings and StarCraft versions (SC2.4.10). Each experiment was carried out using five random seeds. The experimental results are depicted in Figure~\ref{fig:smac_result}.

The results indicate that SIDIFF enhances the performance of the original basic algorithm, as it outperforms QMIX in nearly all tasks. 
COLA, which infers the low-dimensional representation of the global state through contrastive learning, also demonstrates strong performance across various scenarios. Especially in the \emph{3s\_vs\_5z} task, COLA outperforms SIDIFF. 
However, in more challenging tasks such as \emph{8m\_vs\_9m}, \emph{10m\_vs\_11m} and \emph{MMM2}, although COLA performs better than other baselines, it still lags significantly behind SIDIFF. 
The disparity between SIDIFF and COLA suggests that reconstructing the global state within the original state space enhances the agent's decision-making capabilities.\looseness=-1

\subsection{Vectorized Multi-Agent Simulator}

The Vectorized Multi-Agent Simulator (VMAS) is an open-source framework designed for efficient MARL benchmarking. It features a vectorized 2D physics engine implemented in PyTorch~\cite{PaszkeGMLBCKLGA19}, along with a collection of challenging multi-agent scenarios. In these scenarios, all agents can only observe their own positions and information related to their specific goals, which significantly hinders collaborative decision-making among agents. 
VDN and MAPPO are representative methods of value decomposition and multi-agent policy gradient approaches, respectively. To demonstrate that SIDIFF can be easily applied to various algorithms to improve their performance, we compared the performance of the original VDN and MAPPO algorithms to their SIDIFF variants. Figure~\ref{fig:vams_result} illustrates the performance of the original algorithms and their SIDIFF variants across three representative scenarios within VMAS. Appendix~B.2 provides detailed descriptions of these three scenarios.

Since VDN is an off-policy algorithm, it converges more quickly than MAPPO, which is on-policy. For each algorithm, the SIDIFF variant outperforms the original algorithm. Specifically, in the \emph{Wheel} scenario, MAPPO fails to converge. However, agents in MAPPO with SIDIFF can understand the states and intentions of other agents based on the global state, avoiding conflicts between strategies and achieving cooperation more rapidly.

\begin{figure}[!b]
    \vskip -0.15 in
    \centering
    \subfigure[Multi-Agent Battle City.]{
        \includegraphics[width=1.65 in]{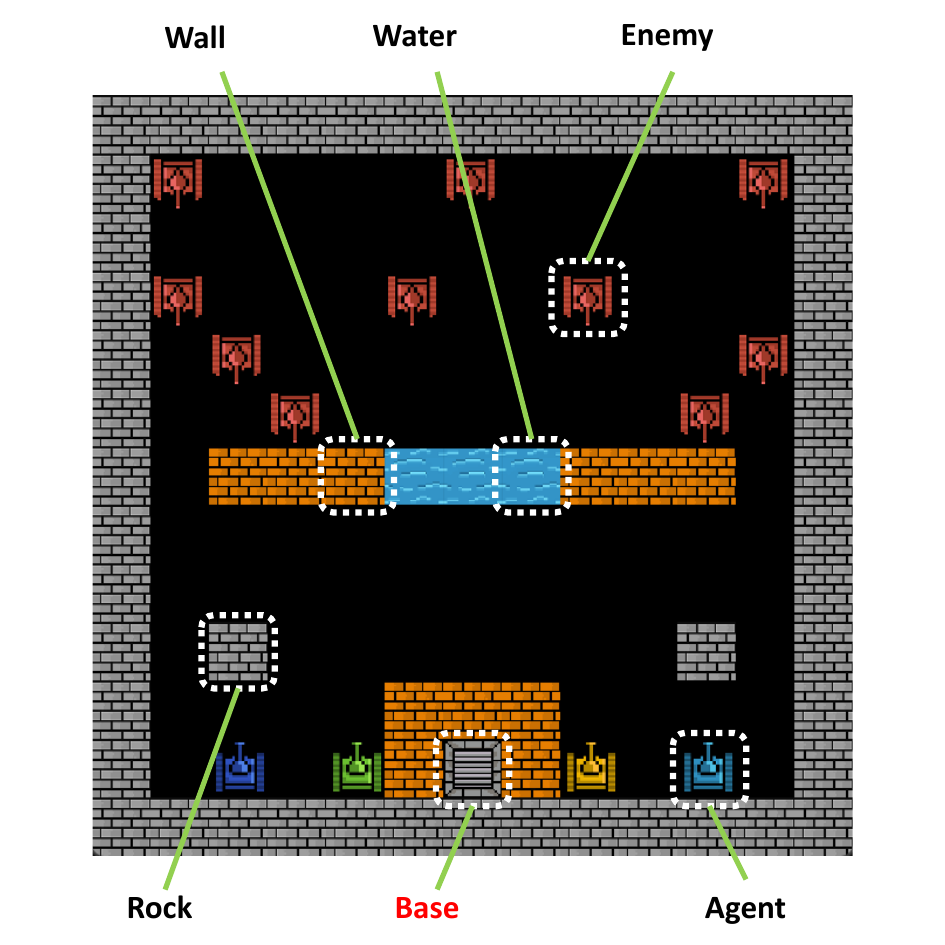}
        \label{fig:mabc_intro}
    }
    \hspace{-0.2 in}
    \subfigure[The \emph{2\_vs\_8} scenario.]{
        \includegraphics[width=1.65 in]{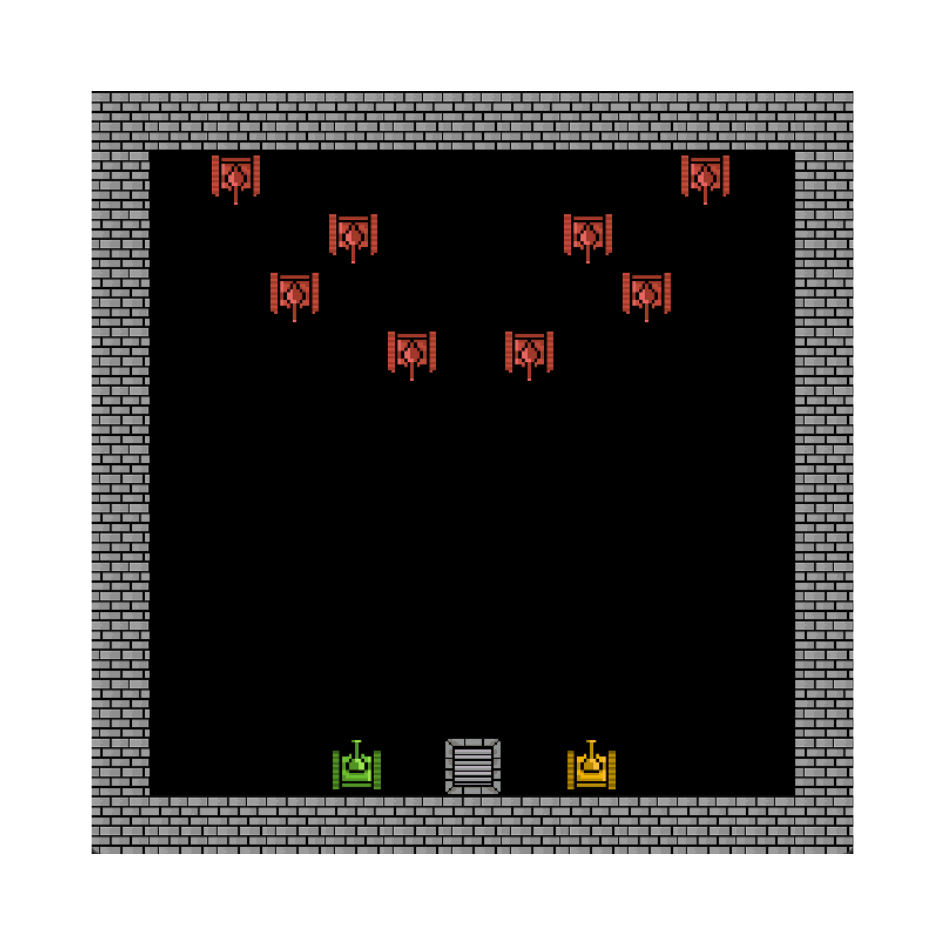}
        \label{fig:mabc_2_vs_8}
    }
    \vskip -0.15 in
    \caption{Screenshots of tasks in Multi-Agent Battle City.}
\end{figure}

\begin{figure}[!t]
    \centering
    \subfigure[The learning curves.]{
        \includegraphics[width=1.65 in]{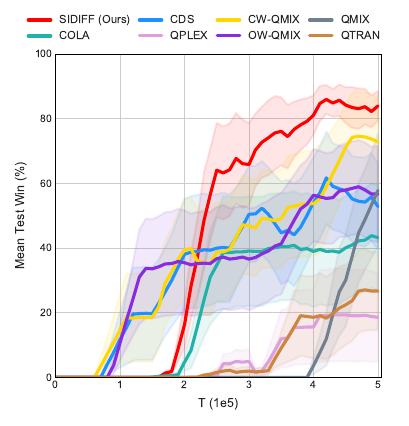}
        \label{fig:mabc_result}
    }
    \hspace{-0.2 in}
    \subfigure[Visualization of policies.]{
        \includegraphics[width=1.65 in]{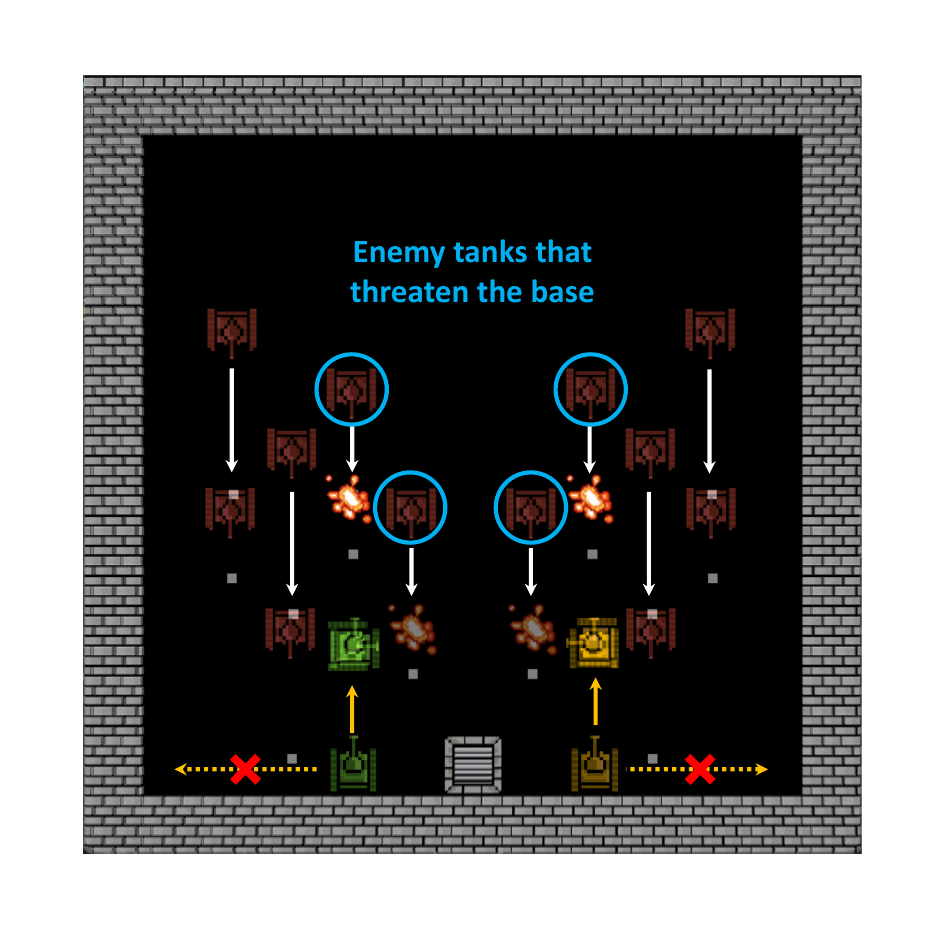}
        \label{fig:mabc_case}
    }
    \vskip -0.15 in
    \caption{Results on MABC benchmark.}
\vspace{-0.2 in}
\end{figure}

\subsection{Case Study: Multi-Agent Battle City}

Battle City is an internationally popular video game from the past century. In this game, a single player controls a tank to protect the base from attacks while destroying enemy tanks, which are controlled by built-in heuristic algorithms. The map contains a variety of terrain, as illustrated in Figure~\ref{fig:mabc_intro}.
We reproduced this game with a simple program and expanded it to enable multiple players to cooperate simultaneously. In MABC, allied tanks can only see information about themselves. 
The behavior of enemy tanks is predetermined, with their specific targeting algorithm being to prioritize the closest allied unit or the base, giving higher priority to the base. 
We designed a \emph{2\_vs\_8} level, with the initial state depicted in Figure~\ref{fig:mabc_2_vs_8}. Detailed information about the environment can be found in Appendix~B.3.

As shown in Figure~\ref{fig:mabc_result}, under conditions of limited information, many algorithms tend to fall into local optima, focusing on destroying enemy tanks while neglecting the protection of the base. The agents controlled by these algorithms initially retreat to a corner to avoid being attacked and then seek opportunities to counterattack. However, once the base is attacked, the game ends and is considered a failure. In contrast, agents controlled by SIDIFF can perceive changes in the global state, enabling them to identify this critical mechanism earlier and avoid falling into the local optimum. Figure~\ref{fig:mabc_case} depicts the specific strategy employed by agents controlled by SIDIFF: They prioritize attacking or drawing away enemy tanks that pose a significant threat to the base.\looseness=-1

\subsection{Ablation Study and Visualization}

\begin{figure}[b]
    \vspace{-0.25 in}
    \centering
    \includegraphics[width=2.8 in]{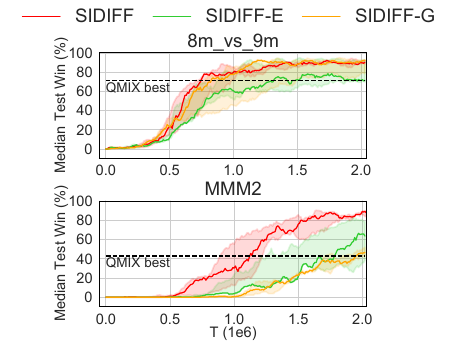}
    \vspace{-0.15 in}
    \caption{Win rates for SIDIFF and ablations.}
    \label{fig:ablation}
\end{figure}

We demonstrate the importance of each module in SIDIFF through two scenarios in SMAC. SIDIFF-E and SIDIFF-G denote versions of SIDIFF without the state generator and state extractor, respectively, while keeping other components unchanged. SIDIFF-E retains the state extractor but lacks the state generator. Therefore, we replace the global state with local observations as input to the state extractor. This approach allows us to verify whether the additional network modules impact the algorithm's performance.
In contrast, SIDIFF-G contains only the state generator. In this variant, we combine the reconstructed global state with local observations and feed them directly into the agent network. By comparing the performance of these two variants with SIDIFF, we can intuitively determine their importance.

In the hard scenario \emph{8m\_vs\_9m} and the super hard scenario \emph{MMM2}, SIDIFF significantly outperforms the other two variants. Furthermore, as shown in Figure~\ref{fig:ablation}, both SIDIFF-E and SIDIFF-G exhibit superior performance in comparison to the original QMIX algorithm. 
This implies that agents require the global state to assist in decision-making and must correctly process this information in order to extract the most useful data.
These experimental results indicate that both the state generator and state extractor components in SIDIFF are essential.

\begin{figure}[t]
    \centering
    \includegraphics[width=3.0 in]{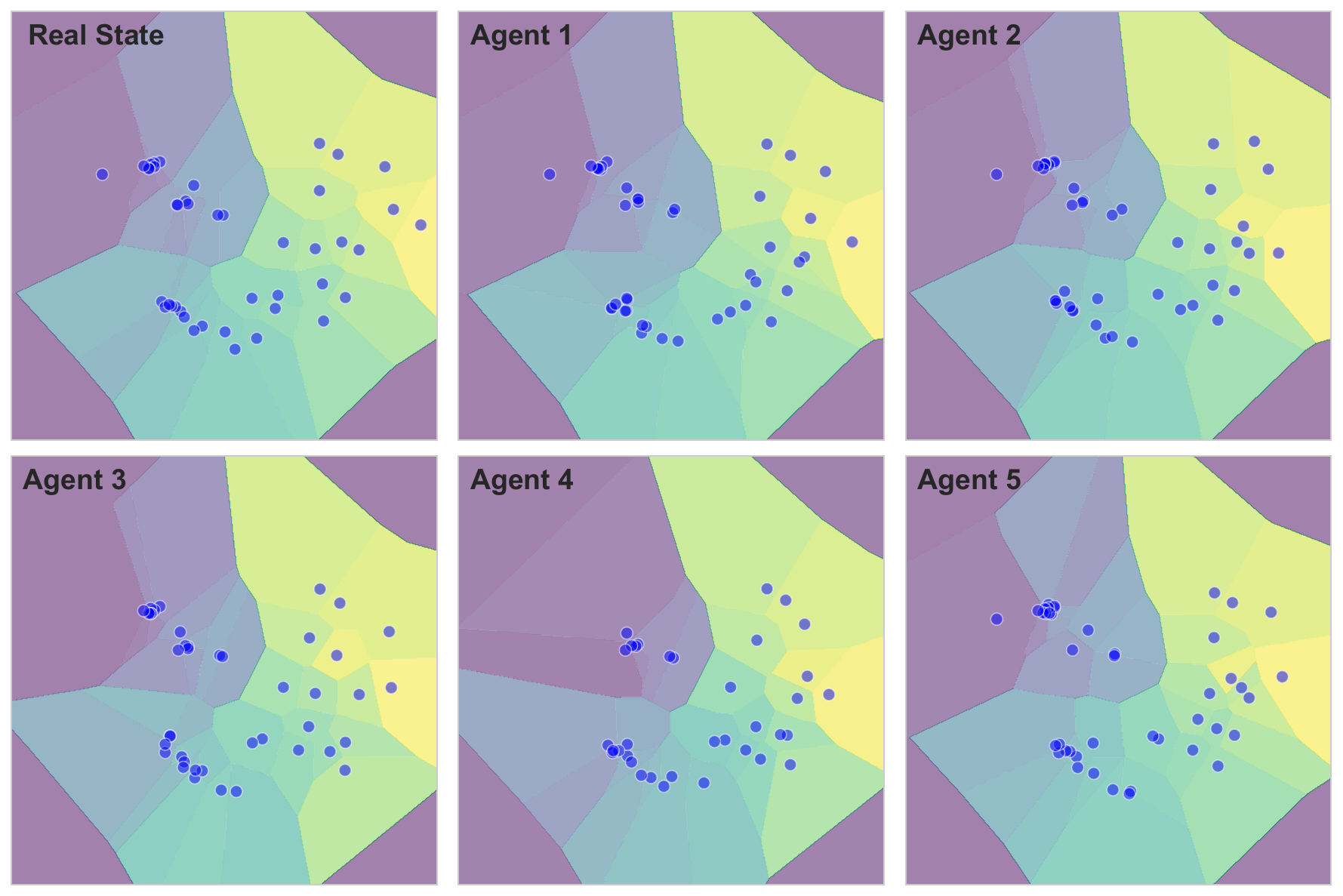}
    \vskip -0.10 in
    \caption{The 2D t-SNE embeddings of the real state and reconstructed states.}
    \vskip -0.19 in
    \label{fig:state_vis}
\end{figure}

To more intuitively demonstrate the agents' ability to infer the global state, we visualized the actual state sequence alongside the states reconstructed by the agents.
For the trajectories of an episode in the \emph{2s3z} scenario, we visualized the t-SNE~\cite{TSNE} embeddings of the reconstructed states and compared them with the real state embeddings.
In Figure~\ref{fig:state_vis}, the first plot shows the real states, while the subsequent plots depict the states inferred by each agent based solely on their local observations. Each point in the diagram represents the embedding of a specific state.
The background color indicates the time step of all states within the filled area. The closer the state appears to the beginning of the episode, the lighter the background color.
From the visualizations, we can observe that the states inferred by all agents closely match the real states, demonstrating that the state generator effectively allows agents to infer the global state based solely on local observations. More experimental results can be found in \textbf{Appendix~C}.


\section{Conclusion and Discussion}

In this paper, we propose SIDIFF to address the issue of agents being unable to access the global state during the decentralized execution phase in Dec-POMDP tasks. 
The state generator in SIDIFF, a diffusion model,  reconstructs the global state in a distributed manner based on the agents' trajectory history. 
The state extractor then processes the reconstructed global state, removing redundant information and extracting decision-relevant information for the agents. 
These two components work together to provide SIDIFF with superior performance in a variety of cooperative multi-agent environments.
In addition, we introduce Multi-Agent Battle City (MABC), a novel multi-agent experimental environment that allows flexible customization of scenarios. 
We believe that MABC is sufficiently simple yet distinctive to incentivize more contributions from the MARL community.

Future work will involve exploring the application of SIDIFF in multi-task multi-agent scenarios. Moreover, replacing the current traditional diffusion model with faster generative diffusion models to improve inference efficiency is a promising research direction.

\bibliography{ref}

\onecolumn
\appendix
\numberwithin{equation}{section}
\numberwithin{figure}{section}
\numberwithin{table}{section}
\renewcommand{\thesection}{{\Alph{section}}}
\renewcommand{\thesubsection}{\Alph{section}.\arabic{subsection}}
\renewcommand{\thesubsubsection}{\Roman{section}.\arabic{subsection}.\arabic{subsubsection}}
\setcounter{secnumdepth}{-1}
\setcounter{secnumdepth}{3}

\section{Implementation Details}

\subsection{Algorithmic Description}

The algorithms for different methods with SIDIFF are summarized in Algorithm~\ref{alg:SIDIFF_VD} and Algorithm~\ref{alg:SIDIFF_PG}. The code for SIDIFF can be found in the supplementary material.

\begin{algorithm}[htbp]
\caption{Value decomposition methods with SIDIFF}
\label{alg:SIDIFF_VD}
\textbf{Hyperparameters}: Patch dimensions $d$, discount factor $\gamma$, exploration coefficient $\epsilon$\\
Initialize the parameters of the agent network, the mixing network and the state extractor\\
Initialize the parameters of the state generator
\begin{algorithmic}[1] 
\FOR{each episode}
\IF{Train}
\STATE Obtain the global state $s_1$ and the local observations $\boldsymbol{z}_1=\{z^1_1, z^2_1,\dots, z^n_1\}$
\FOR{$t \leftarrow 1$ to $T-1$}
\STATE Divide the global state $s$ into patches of dimension $d$
\STATE Calculate the abstracted feature embedding $s_g$ according Eq.~\eqref{eq:abstracted_feat}
\FOR{$a \leftarrow 1$ to $n$}
\STATE Select action $u^a_t$ according to $\epsilon$-greedy policy w.r.t $Q_a(\tau^a_t, s_g)$
\ENDFOR
\STATE Take the joint action $\boldsymbol{u}_t=\{u_t^0, u_t^1,\dots,u_t^n\}$
\STATE Obtain the global reward $r_{t+1}$, the next local observations $\boldsymbol{z}_{t+1}$, and the next state $s_{t+1}$
\ENDFOR
\STATE Store the episode in $\mathcal{D}$
\STATE Sample a batch of episodes $\sim$ Uniform($\mathcal{D}$)
\STATE Update the parameters of the state generator according Eq.~\eqref{eq:diff_loss}
\STATE Update the parameters of the agent network, the mixing network and the state extractor according TD loss
\STATE Replace target parameters every $M$ episodes
\ENDIF
\IF{Evaluate}
\FOR{$t \leftarrow 1$ to $T-1$}
\FOR{$a \leftarrow 1$ to $n$}
\STATE Obtain the local observation $z^a_t$
\STATE Reconstruct the global state $\hat{s}_t^a$ according Eq.~\eqref{eq:denoise}
\STATE Divide the reconstructed global state $\hat{s}_t^a$ into patches of dimension $d$
\STATE Calculate the abstracted feature embedding $s_g$ according Eq.~\eqref{eq:abstracted_feat}
\STATE Select action $u^a_t$ according to greedy policy w.r.t $Q_a(\tau^a_t, s_g)$
\ENDFOR
\ENDFOR
\ENDIF
\ENDFOR
\end{algorithmic}
\end{algorithm}

\subsection{Implementation Details and Hyperparameters}

In practical implementation, to ensure algorithm training stability, we typically update the state generator only before evaluation. This approach ensures the effectiveness of the state generator in inferring states during evaluation and also reduces training time. When optimizing the state generator, we typically use the trajectories from the most recent $L$ episodes as the training dataset to avoid significant differences between the agent's current policy and the historical transition data. By default, $L = 5$. Furthermore, the historical trajectories, which serve as the basis for reconstructing the global state, usually include observations from the past four steps and the current step.

The abstracted feature embedding $s_g$ output by the state extractor is eventually combined with the agent's local observations and fed into either the agent network (in value decomposition methods) or the actor network (in policy gradient methods). Furthermore, to mitigate the compounding error caused by minor deviations in the reconstructed state $\hat{s}$, we feed $s_g$ into a fully connected layer following the RNN layer.

Table~\ref{table:hyperparameters} shows the hyperparameters used for various algorithms. The hyperparameters in variants with SIDIFF are consistent with those in the original algorithms. Furthermore, we ensure that the hyperparameters for other baselines match those specified in their official code. All experiments in this paper are run on Nvidia GeForce RTX 3090 graphics cards and Intel(R) Xeon(R) Platinum 8280 CPU. As a result, the solid lines represent the median win rates and the 25-75\% percentiles are shaded.

\begin{algorithm}[H]
\caption{Policy gradient methods with SIDIFF}
\label{alg:SIDIFF_PG}
\textbf{Hyperparameters}: Patch dimensions $d$, discount factor $\gamma$\\
Initialize the parameters of the actor, the critic and the state extractor\\
Initialize the parameters of the state generator
\begin{algorithmic}[1] 
\FOR{each episode}
\IF{Train}
\STATE Obtain the global state $s$ and local observations $\boldsymbol{z}_1=\{z^1_1, z^2_1,\dots, z^n_1\}$
\FOR{$t \leftarrow 1$ to $T-1$}
\FOR{$a \leftarrow 1$ to $n$}
\STATE Calculate the abstracted feature embedding $s_g$ according Eq.~\eqref{eq:abstracted_feat}
\STATE Sample the action from the distribution w.r.t $u^a_t=\pi_{a}(z^a_t, s_g)$
\ENDFOR
\STATE Take the joint action $\boldsymbol{u}_t=\{u_t^1, u_t^2,\dots,u_t^n\}$
\STATE Obtain the global reward $r_{t+1}$, the next local observations $\boldsymbol{z}_{t+1}$, and the next state $s_{t+1}$
\STATE Store the episode in replay buffer $\mathcal{D}$
\FOR{$a \leftarrow 1$ to $n$}
\STATE Sample a batch of episodes $\sim$ Uniform($\mathcal{D}$)
\STATE Fit value function by regression on mean-squared error and update critic
\STATE Update actor according to policy gradient methods (such as maximizing the PPO-Clip objective for MAPPO)
\ENDFOR
\STATE Update target parameters by polyak averaging
\ENDFOR
\ENDIF
\IF{Evaluate}
\FOR{$t \leftarrow 1$ to $T-1$}
\FOR{$a \leftarrow 1$ to $n$}
\STATE Obtain the local observation $z^a_t$
\STATE Reconstruct the global state $\hat{s}_t^a$ according Eq.~\eqref{eq:denoise}
\STATE Divide the reconstructed global state $\hat{s}_t^a$ into patches of dimension $d$
\STATE Calculate the abstracted feature embedding $s_g$ according Eq.~\eqref{eq:abstracted_feat}
\STATE Sample the action from the distribution w.r.t $u^a_t=\pi_{a}(z^a_t, s_g)$
\ENDFOR
\ENDFOR
\ENDIF
\ENDFOR
\end{algorithmic}
\end{algorithm}

\begin{table}[H]
\centering
\begin{tabular}{lll}
\hline
\textbf{Algorithm}                 & \textbf{Description}                                        & \textbf{Value} \\ \hline
\multirow{7}{*}{State Generator}   & Type of optimizer                                           & Adam           \\
                                   & Learning rate                                               & 0.0005         \\
                                   & Number of last episodes used for the update $L$             & 10             \\
                                   & Batch size                                                  & 1 (episode)    \\
                                   & Number of epochs                                            & 1000           \\
                                   & Diffusion process duration $K$                              & 20             \\
                                   &  probability for unconditional class identifier $p$
                                        & 0.001          \\ \hline
\multirow{2}{*}{State Extractor}   & Dimension of each patch $d$                                 & 5              \\
                                   & Number of attention heads $H$                               & 4              \\ \hline
\end{tabular}
\caption{Hyperparameter settings.}
\label{table:hyperparameters}
\end{table}


\section{Environment Details}

\subsection{StarCraft Multi-Agent Challenge (SMAC)}

The StarCraft Multi-Agent Challenge (SMAC) is a benchmark environment designed to test and develop MARL algorithms within the complex setting of the real-time strategy game StarCraft II.
SMAC focuses on micromanagement tasks, where individual units are controlled by independent learning agents that must coordinate their actions based solely on local observations. 
Key features of SMAC include a diverse set of combat scenarios that challenge agents to handle high-dimensional inputs, partial observability, and the need for sophisticated cooperation to achieve victory. 
The environment supports centralized training with decentralized execution, providing additional state information during training to facilitate algorithm development. SMAC also offers standardized evaluation metrics and best practices, promoting rigorous experimental methodologies in the MARL field. Table~\ref{table:SMAC_scenario} presents the details of several representative scenarios in SMAC.

\begin{table}[h]
\centering
\begin{tabular}{lccc}
\hline
\textbf{Name}  & \textbf{Ally Units}                                                         & \textbf{Enemy Units}                                                        & \textbf{Type}                                                                          \\ \hline
2s3z           & \begin{tabular}[c]{@{}c@{}}2 Stalkers\\ 3 Zealots\end{tabular}              & \begin{tabular}[c]{@{}c@{}}2 Stalkers\\ 3 Zealots\end{tabular}              & \begin{tabular}[c]{@{}c@{}}Heterogeneous\\ Symmetric\end{tabular}                     \\ \hline
3s5z           & \begin{tabular}[c]{@{}c@{}}3 Stalkers\\ 5 Zealots\end{tabular}              & \begin{tabular}[c]{@{}c@{}}3 Stalkers\\ 5 Zealots\end{tabular}              & \begin{tabular}[c]{@{}c@{}}Heterogeneous\\ Symmetric\end{tabular}                     \\ \hline
1c3s5z         & \begin{tabular}[c]{@{}c@{}}1 Colossus\\ 3 Stalkers\\ 5 Zealots\end{tabular} & \begin{tabular}[c]{@{}c@{}}1 Colossus\\ 3 Stalkers\\ 5 Zealots\end{tabular} & \begin{tabular}[c]{@{}c@{}}Heterogeneous\\ Symmetric\end{tabular}                     \\ \hline
3s\_vs\_5z     & 3 Stalkers                                                                  & 5 Zealots                                                                   & \begin{tabular}[c]{@{}c@{}}Homogeneous\\ Asymmetric\end{tabular}                      \\ \hline
2c\_vs\_64zg   & 2 Colossi                                                                   & 64 Zerglings                                                                & \begin{tabular}[c]{@{}c@{}}Homogeneous\\ Asymmetric\end{tabular}                      \\ \hline
5m\_vs\_6m     & 5 Marines                                                                   & 6 Marines                                                                   & \begin{tabular}[c]{@{}c@{}}Homogeneous\\ Asymmetric\end{tabular}                      \\ \hline
8m\_vs\_9m     & 8 Marines                                                                   & 9 Marines                                                                   & \begin{tabular}[c]{@{}c@{}}Homogeneous\\ Asymmetric\end{tabular}                      \\ \hline
10m\_vs\_11m   & 10 Marines                                                                  & 11 Marines                                                                   & \begin{tabular}[c]{@{}c@{}}Homogeneous\\ Asymmetric\end{tabular}                     \\ \hline
MMM2           & \begin{tabular}[c]{@{}c@{}}1 Medivac\\ 2 Marauders\\ 7 Marines\end{tabular} & \begin{tabular}[c]{@{}c@{}}1 Medivac\\ 3 Marauder\\ 8 Marines\end{tabular}  & \begin{tabular}[c]{@{}c@{}}Heterogeneous\\ Asymmetric\\ Macro tactics\end{tabular}    \\ \hline
\end{tabular}
\caption{Maps in different scenarios.}
\label{table:SMAC_scenario}
\end{table}

\subsection{Vectorized Multi-Agent Simulator (VMAS)}

The Vectorized Multi-Agent Simulator (VMAS) is an open-source framework designed for efficient MARL benchmarking. It features a vectorized 2D physics engine developed in PyTorch and includes a suite of challenging multi-robot scenarios that can be easily expanded through a user-friendly, modular interface. VMAS is distinguished by its capability for parallel simulation on accelerated hardware, offering significant speed advantages over traditional simulators. This framework is aimed at fostering advancements in collective robot learning. At the beginning of each episode, we fix the position of each agent to reduce the learning complexity.
Each scenario we tested is described in detail below.

\begin{figure}[H]
    \centering
    \subfigure[Dispersion.]{
    \includegraphics[width=1.4 in]{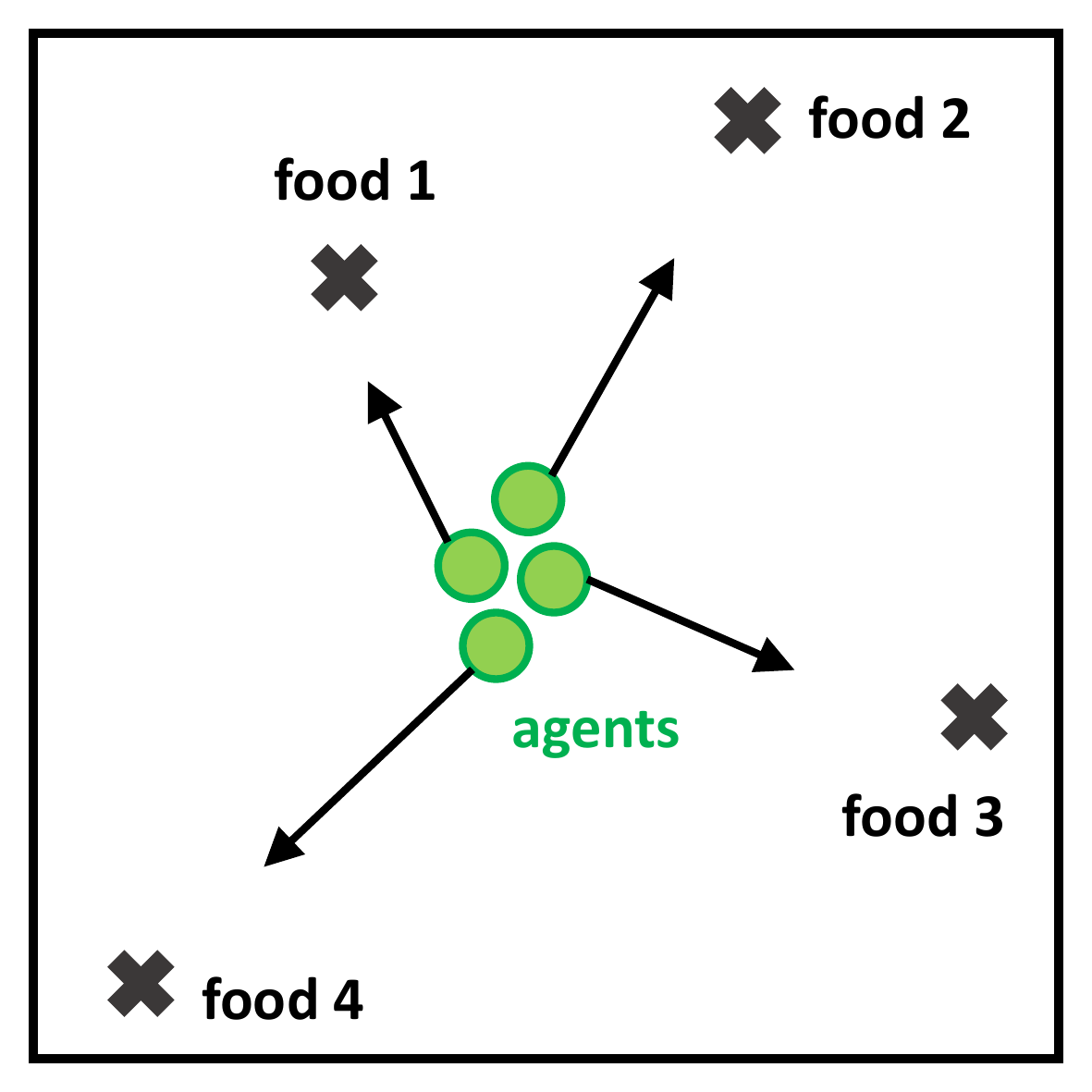}
    \label{fig:dispersion}
    }
    \hspace{0.1 in}
    \subfigure[Passage.]{
    \includegraphics[width=1.4 in]{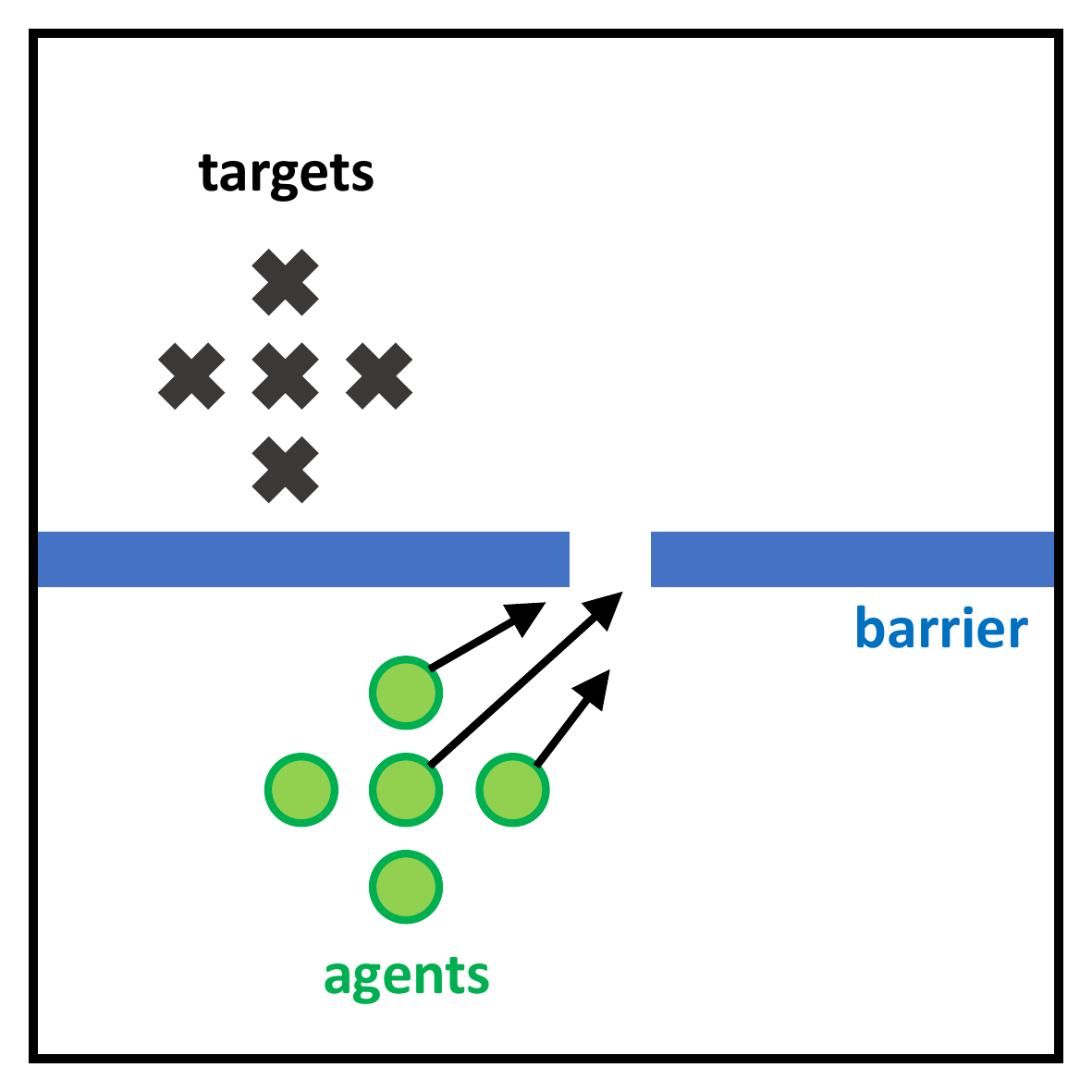}
    \label{fig:passage}
    }
    \hspace{0.1 in}
    \subfigure[Wheel.]{
    \includegraphics[width=1.4 in]{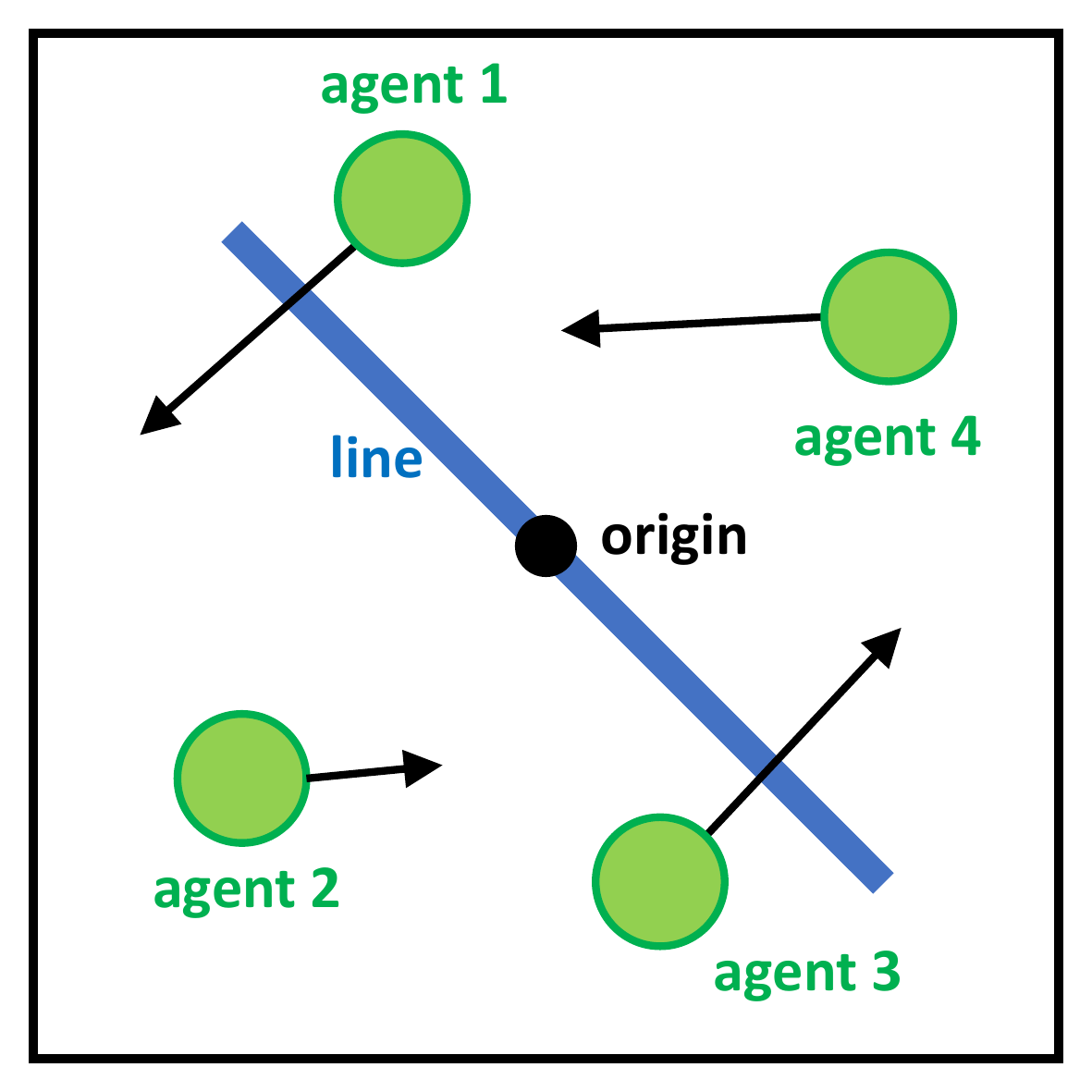}
    \label{fig:wheel}
    }
\caption{Illustrations of the Vectorized Multi-Agent Simulator.}
\label{fig:VMAS_scenarios}
\end{figure}

\noindent{\textbf{Dispersion.}}\quad The agents must collaboratively consume all available food particles scattered throughout the environment. Starting from a common initial position, agents must navigate to and consume different food particles.

\noindent{\textbf{Passage.}}\quad A group of five agents starts in a cross formation and must replicate this formation on the opposite side of a barrier. The barrier has one passage, and agents are penalized for colliding with either each other or the barrier. The agents must navigate through the barrier's passages without collisions, which involves complex strategies to coordinate their movements effectively.

\noindent{\textbf{Wheel.}}\quad A group of agents is tasked with collectively rotating a line anchored at its origin. This line has customizable mass and length, and the team's objective is to achieve a specific angular velocity. The challenge arises from the fact that a single agent cannot move a high-mass line alone, necessitating coordinated efforts among all agents. They must organize themselves so that some agents increase the line's velocity while others reduce it, ultimately achieving the desired state of motion.

\subsection{Multi-Agent Battle City (MABC)}

The Multi-Agent Battle City (MABC) environment is a Python-based platform intended for MARL research. It is inspired by the classic game \emph{Battle City} and includes a number of features such as support for various terrains, multiple players (up to four by default), and customizable maps. The platform is designed to be easily extended. This testbed is distinguished by its partially observable setting and sparse reward system, both of which necessitate complex strategic planning to achieve victory.

At each time step, agents receive local observations drawn within their field of view. Agents can only observe other agents if they are alive and located within the sight range. The feature vector observed by each agent contains the following attributes for allied and enemy units within the sight range: \texttt{relative x}, \texttt{relative y}, \texttt{orientation}, and \texttt{remaining cooldown time for missile firing}. Additionally, agents can obtain their own absolute position, orientation, and cooldown time. The global state, which is only available to agents during centralized training, contains information about all units on the map. All features, both in the state and in the observations of individual agents, are normalized by their maximum values. The sight range of all agents can be adjusted according to the needs. Moreover, MABC provides both raw pixel data and low-dimensional feature information for states and observations.

In MABC, all agents (tanks) share the same action space. The discrete set of actions that agents are allowed to take consists of \texttt{move [left, right, up, down]}, \texttt{fire}, and \texttt{no-op}. The overall goal is to maximize the win rate for each battle scenario. The default setting uses a shaped reward, which produces a reward based on the hit-point damage dealt and enemy units killed, together with a special bonus for winning the battle. We also provide a sparse reward option, in which the reward is +1 for winning and 0 for losing an episode. 

In contrast to the SMAC environment, the allied agents in MABC face the dual challenge of not only eliminating enemy units but also defending a base located at the bottom of the map from enemy attacks. If the base is attacked, the episode ends and is considered a failure. This requires the agents to develop strategies with a global perspective.

\begin{figure}[H]
    \centering
    \includegraphics[width=4.5 in]{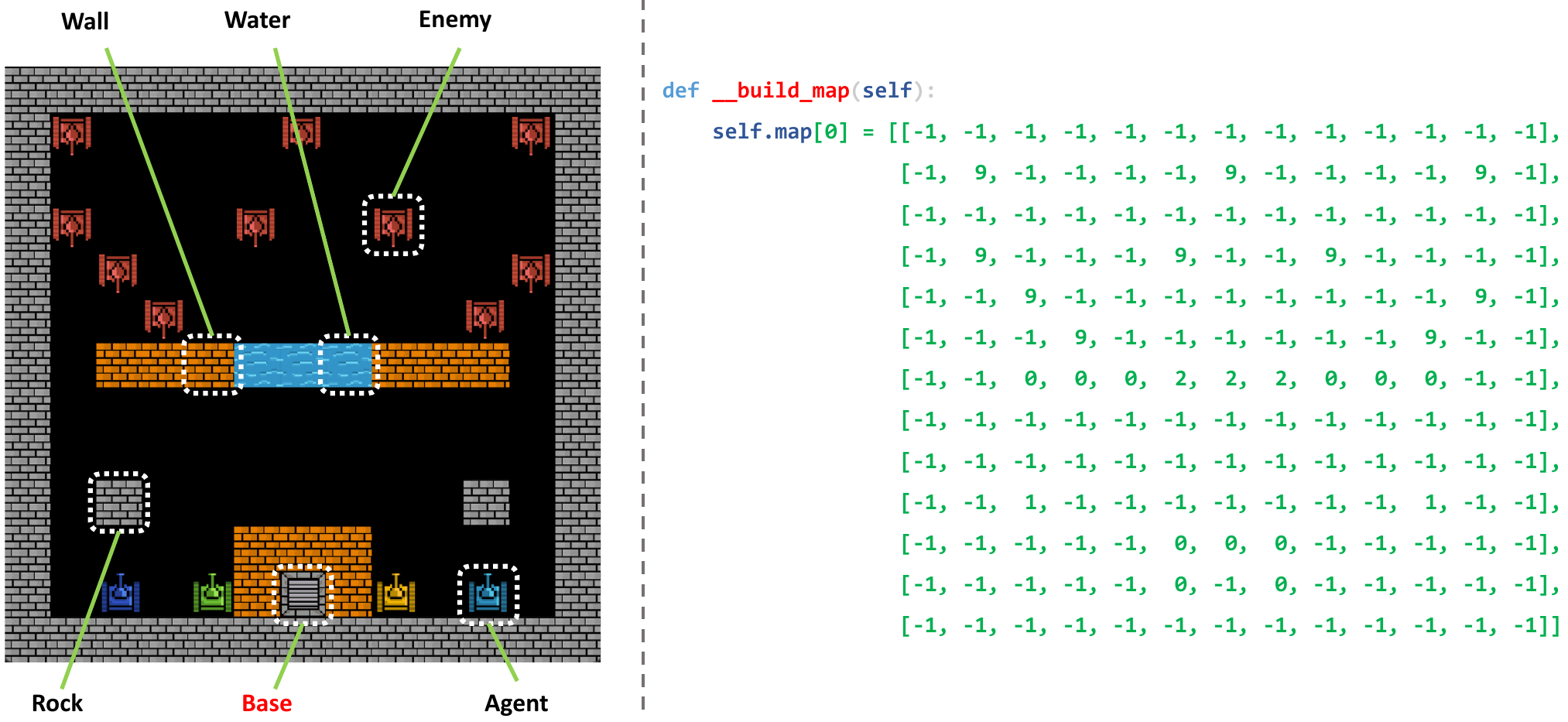}
    \caption{In MABC, the scenario terrain can be easily configured through code.}
    \label{fig:terrain}
\end{figure}

MABC is a highly customizable environment. As illustrated in Figure~\ref{fig:terrain}, the terrain and the spawn points of enemy units in each scenario can be quickly configured by editing the code. In addition, the strategies of enemy units can be flexibly adjusted. We believe that this simple yet unique environment has the potential to attract the attention of both the MARL and large language model (LLM) communities.


\section{Additional Experiments}

\subsection{Additional Ablation Experiments}

\begin{figure*}[t]
    \centering
    \subfigure{
    \includegraphics[width=0.45\linewidth]{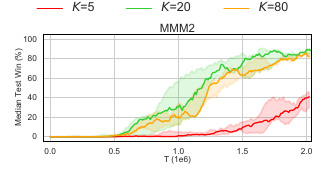}
    }
    \hspace{0.2 in}
    \subfigure{
    \includegraphics[width=0.45\linewidth]{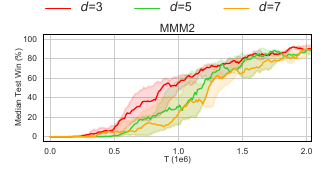}
    }
    \caption{Influence of the $K$ and $d$ for SIDIFF.}
    \label{fig:addition_ablation}
\end{figure*}

We demonstrate the impact of various hyperparameter settings on the performance of SIDIFF in Figure~\ref{fig:addition_ablation}. Specifically, we examine how different values of the diffusion process duration $K$ and the patch dimension $d$ affect the performance. As $K$ increases, the performance of SIDIFF initially improves and then declines. We attribute this to the fact that when $K$ is low, the state generator cannot accurately reconstruct the global state with fewer iterations. However, when $K$ is large, the convergence speed of the state generator is slower than the reinforcement learning process, given the fixed number of training iterations for the state generator. Considering both time and training costs, the value of $K$ should be chosen to be relatively optimal. Additionally, regarding the dimension of each patch $d$, we find that smaller values of $d$ lead to better performance of SIDIFF, indicating that finer-grained division of the state can capture more critical information. However, smaller $d$ values result in higher computational costs, thus necessitating a trade-off between computational cost and performance.

\subsection{Additional Visualization}

\begin{figure}[H]
    \centering
    \includegraphics[width=6.8 in]{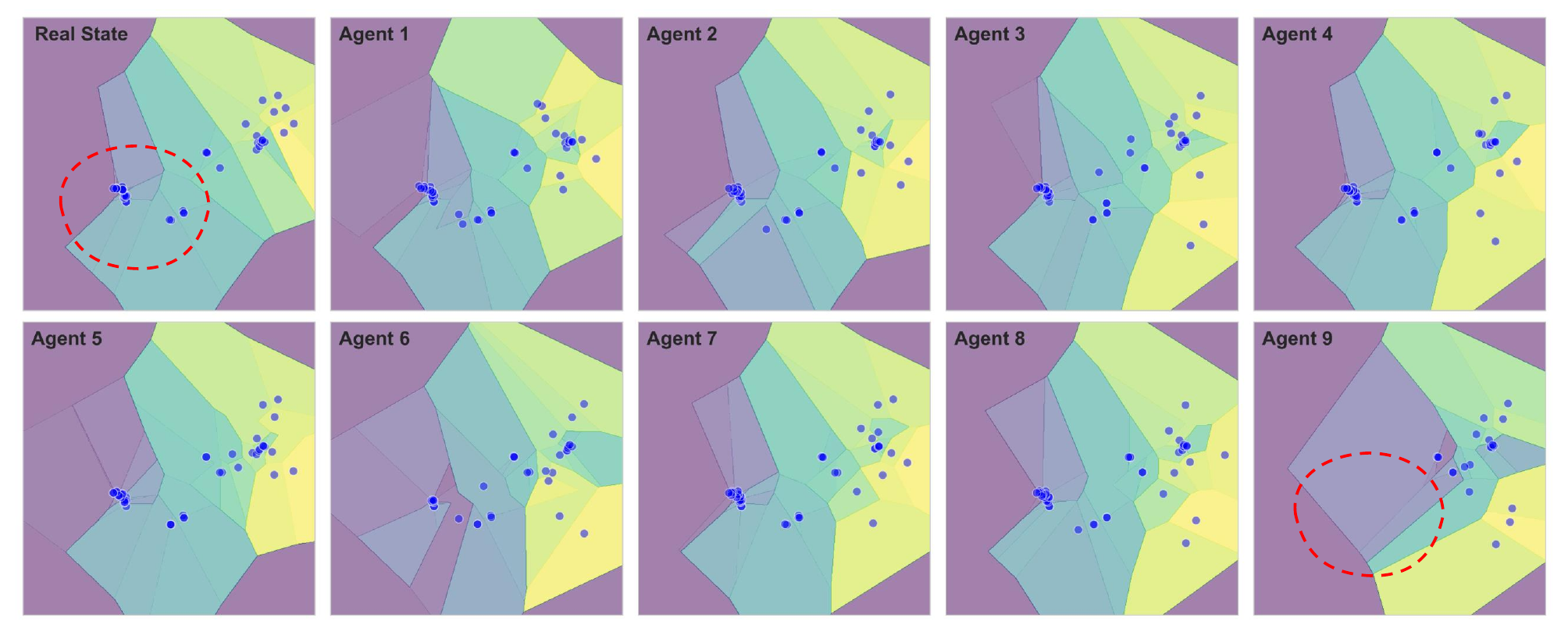}
    \vskip -0.10 in
    \caption{The 2D t-SNE embeddings of the real state and reconstructed states in \emph{MMM2} scenario.}
    \vskip -0.19 in
    \label{fig:vis_appendix}
\end{figure}

To further confirm the state generator's ability to reconstruct states, we also present the 2D t-SNE embeddings of the real states and reconstructed states in the super hard scenario. As shown in Figure~\ref{fig:vis_appendix}, even in tasks with a complex state space, the reconstructed states produced by the state generator are generally consistent with the real states. The discrepancies highlighted by the red dashed circles are due to the agent's death before the end of the episode.

\end{document}